\def\bold#1{\setbox0=\hbox{$#1$}%
     \kern-.025em\copy0\kern-\wd0
     \kern.05em\copy0\kern-\wd0
     \kern-.025em\raise.0433em\box0 }

\def\slash#1{\setbox0=\hbox{$#1$}#1\hskip-\wd0\dimen0=5pt\advance
       \dimen0 by-\ht0\advance\dimen0 by\dp0\lower0.5\dimen0\hbox
         to\wd0{\hss\sl/\/\hss}}

\documentstyle[12pt]{article}

\newlength{\dinwidth}
\newlength{\dinmargin}
\newcommand{\resection}[1]{\setcounter{equation}{0}\section{#1}}

\begin{document}

\def\lq{\left [}
\def\rq{\right ]}
\def\LL{{\cal L}}
\def\VV{{\cal V}}
\def\AA{{\cal A}}

\newcommand{\be}{\begin{equation}}
\newcommand{\ee}{\end{equation}}
\newcommand{\bea}{\begin{eqnarray}}
\newcommand{\eea}{\end{eqnarray}}
\newcommand{\nn}{\nonumber}
\newcommand{\dd}{\displaystyle}

\thispagestyle{empty}
\vspace*{4cm}
\begin{center}
  \begin{Large}
  \begin{bf}
ESTIMATES WITH AN EFFECTIVE CHIRAL LAGRANGIAN FOR HEAVY MESONS$^*$\\
  \end{bf}
  \end{Large}
  \vspace{5mm}
  \begin{large}
R. Casalbuoni\\
  \end{large}
Dipartimento di Fisica, Univ. di Firenze\\
I.N.F.N., Sezione di Firenze\\
  \vspace{5mm}
  \begin{large}
A. Deandrea, N. Di Bartolomeo and R. Gatto\\
  \end{large}
D\'epartement de Physique Th\'eorique, Univ. de Gen\`eve\\
  \vspace{5mm}
  \begin{large}
F. Feruglio\\
  \end{large}
Dipartimento di Fisica, Univ.
di Padova\\
I.N.F.N., Sezione di Padova\\
  \vspace{5mm}
  \begin{large}
G. Nardulli\\
  \end{large}
Dipartimento di Fisica, Univ.
di Bari\\
I.N.F.N., Sezione di Bari\\
  \vspace{5mm}
\end{center}
  \vspace{2cm}
\begin{center}
UGVA-DPT 1992/07-779\\\
BARI-TH/92-117\\\
Revised version, September 1992
\end{center}
\vspace{1cm}
\noindent
$^*$ Partially supported by the Swiss National Foundation
\newpage
\thispagestyle{empty}
\begin{quotation}
\vspace*{5cm}
\begin{center}
  \begin{Large}
  \begin{bf}
  ABSTRACT
  \end{bf}
  \end{Large}
\end{center}
  \vspace{5mm}
\noindent
On the basis of an effective lagrangian incorporating approximate
chiral symmetry and heavy-quark spin and flavor symmetries, and by use of
information on leptonic decays, we estimate the effective
$D^\star D\pi$ coupling.
\end{quotation}

\newpage
\setcounter{page}{1}
\resection{Introduction}

Much work has been recently devoted to formulate a heavy-quark
effective theory, for hadrons containing a heavy quark, incorporating
the heavy-quark flavour symmetry and the heavy-quark spin symmetry
that are expected to hold for infinite heavy-quark mass, and
estimate deviations from such a limit \cite{zero}. We recall that another
long-standing attack to the complexity of the QCD dynamics has been the
effective chiral approach, based on the approximate chiral invariance
of the light QCD sector for small masses of the light quarks. It is
important to look for possibilities of using at the same time both
systems of symmetries (heavy quark and chiral) in appropriate kinematical
situations where they both, separately and acting on the different fields,
may be expected to have some approximate validity. The quantitative
analysis of the involved approximations will undoubtedly require efforts
and hard calculations of corrections terms. We shall attempt here some
first estimates, openly based on a certain optimism, considering
to come back to more accurate evaluations of the approximations involved
when some theoretical issues will be solved and possibly more data
accumulated.

In a recent paper \cite{Wise} (see also \cite{Yan},
\cite{Lee}) an effective
chiral symmetric lagrangian describing low momentum interactions of
heavy mesons with the pseudo Goldstone bosons of the $0^-$ octet has been
constructed. The
lagrangian is as follows
\bea
\LL &=&\frac{f^2}{8}Tr\left[\partial^\mu\Sigma\partial_\mu
\Sigma^\dagger\right]+\lambda_0Tr[m_q\Sigma+m_q\Sigma^\dagger]\nn\\
& &-i Tr[{\bar H}_av_\mu\partial^\mu H_a]+\frac{i}{2}Tr[{\bar H}_a v^\mu
H_b(\xi^\dagger\partial_\mu\xi+\xi\partial_\mu\xi^\dagger)_{ba}]\nn\\
& &+i\frac{g}{2}Tr[{\bar H}_a H_b\gamma^\mu\gamma_5
(\xi^\dagger\partial_\mu\xi-\xi\partial_\mu\xi^\dagger)_{ba}]\nn\\
& &+\lambda_1Tr[{\bar H}_a H_b(\xi m_q\xi+\xi^\dagger m_q\xi^\dagger)_{ba}]
\nn\\
& &+\lambda_1^\prime Tr[{\bar H}_a H_a(m_q\Sigma+m_q\Sigma^\dagger)_{bb}]
\nn\\
& &+\frac{\lambda_2}{m_Q}Tr[{\bar H}_a\sigma_{\mu\nu} H_a\sigma^{\mu\nu}]+
\cdots
\eea

Here the fields $H_a$ describe the mesons ${\bar q}_a Q$ made up by the heavy
quark $Q$ and the light anti-quark ${\bar q}_a$ ($a=1,2,3$). The lagrangian
is assumed to describe the limit $m_Q\to\infty$, where, as shown in refs.
\cite{Isgur}, \cite{Voloshin}, QCD has an additional spin flavour symmetry and
a heavy quark effective theory (HQET) can be developed \cite{Georgi}. In
eq. (1.1) $f=132~MeV$, $m_q$ is the diagonal light quark mass matrix; $H_a$ is
a $4\times 4$ matrix that contains the pseudoscalar and
vector meson fields $P_a$ and $P_a^\mu$
\bea
H_a &=& \frac{(1+\slash v)}{2}[P_{a\mu}^\star\gamma^\mu-P_a\gamma_5]\nn\\
{\bar H}_a &=& \gamma_0 H_a^\dagger\gamma_0
\eea
with $P_{a\mu}^\star v^\mu=0$.
These fields have dimension $3/2$ since factors of
$\sqrt{m_P}$ have been absorbed in their definition. We also note that the
meson fields depend on the meson four-velocity $v_\mu$; each term in eq. (1.1)
which is bilinear in $H_a$ is diagonal in $v$, according to the velocity
superselection rule \cite{Georgi} and a sum over velocities is understood.
Finally $H_a$ also depends on the heavy flavour $Q$.

As for light mesons, the matrix $\Sigma$ contains the pseudo Goldstone bosons
in the form
\be
\Sigma=\xi^2=\exp{\frac{2iM}{f}}
\ee
where $M$ is the usual $3\times 3$ matrix of the $0^-$ pseudoscalar octet.

Besides Lorentz invariance and discrete symmetries, the strong interaction
lagrangian (1.1) possesses, in the limit $m_Q\to\infty$, $m_q\to 0$, a
$SU(2)$ spin symmetry, a heavy flavour symmetry and a
$SU(3)_L\otimes SU(3)_R$ symmetry. Under the chiral symmetry the fields
transform as follows
\bea
\Sigma &\to & L\Sigma R^\dagger\nn\\
\xi &\to & L\xi U^\dagger=U\xi R^\dagger\nn\\
H &\to & H U^\dagger
\eea
where $U$ is, in general, a non-linear function of the fields and the
matrices $L\in SU(3)_L$ and $R\in SU(3)_R$. Under the
$SU(2)$ spin symmetry
\be
H_a\to S H_a
\ee
with $S\in SU(2)$.

Finally in eq. (1.1) terms with additional derivatives are
omitted, since one aims at describing only the low momentum interactions of
the light mesons; the ellipsis in (1.1) denotes these terms as well as higher
order mass corrections ($1/m_Q$, etc.).

\resection{Semileptonic decays}

Weak interactions between light and heavy mesons can be described by the
weak current \cite{Wise}
\be
L_a^\mu=i\frac{\alpha}{2}Tr[\gamma^\mu(1-\gamma_5)H_b\xi^\dagger_{ba}]+\cdots
\ee
where again the ellipsis denotes terms vanishing in the limit $m_q\to 0$,
$m_Q\to\infty$ or terms with derivatives. This current has the same
transformation properties of the quark weak current ${\bar q}_a\gamma^\mu
(1-\gamma_5)Q$ under the chiral group, i.e. $({\bar 3}_L, 1_R)$. The
constant $\alpha$ can be obtained by considering the matrix element of
$L_a^\mu$ between the meson state and the vacuum:
\be
\langle 0|{\bar q}\gamma^\mu\gamma_5 Q| P\rangle=if_Pm_Pv^\mu
\ee
In the limit $m_Q\to\infty$ one has the result:
\be
\alpha=f_P\sqrt{m_P}
\ee
and $\alpha$ has a logarithmic dependence on the heavy quark
masses. The scaling law $m_P^{-1/2}$ for $f_P$ predicted by (2.3) has
relevant ${\cal O}(1/m_Q)$ corrections at the charm mass, as shown by QCD
sum rules \cite{Dominguez} and lattice QCD calculations \cite{Allton}. We
take into account these ${\cal O}(1/m_Q)$ terms
since they represent a well defined set of mass
corrections and we assume for $f_D$ and $f_B$
the central values obtained by the QCD sum rule analyses \cite{Dominguez}
(recent lattice QCD analyses favour values of $f_B$
slightly larger \cite{Allton}):
\be
f_D \approx  200~MeV
\ee
\be
f_B \approx  180~MeV
\ee

In this letter we
will present an estimate of the constant $g$ in (1.1).
We shall see that this can be done by using the experimental data
$\Gamma(D^0\to\pi^- e^+\nu_e)=(3.9^{+2.3}_{-1.2})\times 10^{-3}/\tau_{D^0}$
\cite{PDG}, together with the
results (2.4) and
(2.5) and reasonable assumptions on semileptonic form factors.

Let us first consider the semileptonic  $D\to\pi$ decays. With the usual
definition
\be
\langle \pi^-(p_\pi)|{\bar d}\gamma^\mu(1-\gamma_5)c|D^0(p_D)\rangle=
f_+(q^2)(p_D+p_\pi)^\mu+f_-(q^2)(p_D-p_\pi)^\mu
\ee
it follows from (1.1) and (2.1) that the form factors $f_+$ and $f_-$
have the following contributions at $q^2=q^2_{\rm max}=(m_D-m_\pi)^2$:
\be
f_+(q^2_{\rm max})=-\frac{f_D}{2f}\left(1+g\frac{m_D-m_\pi}{\Delta+m_\pi}
\right)
\ee
\be
f_-(q^2_{\rm max})=-\frac{f_D}{2f}\left(1-g\frac{m_D+m_\pi}{\Delta+m_\pi}
\right)
\ee
where $\Delta=m_{D^\star}-m_D=145~MeV$ is the mass difference between the
$1^-$ and $0^-$ low-lying charmed mesons. The value $f_+(0)$ can be
tentatively
obtained, by  assuming for the form factor the $q^2$ dependence suggested by
single pole dominated dispersion relations:
\be
f_+(q^2)=m_1^2\frac{f_+(0)}{m_1^2-q^2}
\ee
where $m_1$ is the mass of the vector meson state with the same
quantum numbers of the current ${\bar q}_a\gamma^\mu Q$ (the
$D^\star(2010)$ state in the present case). From this it follows
\be
f_+(0)=-\frac{f_D}{2f}\left(1+g\frac{m_D-m_\pi}{\Delta+m_\pi}\right)
\frac{(\Delta+m_\pi)(2m_D+\Delta-m_\pi)}{m_{D^\star}^2}
\ee
{}From experimental data on $D^0\to\pi^- e^+\nu_e$ decay \cite{PDG} and
eqs. (2.4) and (2.5),
we get two possible values for $g$ according to the sign of
$f_+(0)$:
\bea
g &=& 0.48\pm 0.15~~~~~~(f_+(0)<0)\nn\\
g &=& -0.80\pm 0.15~~~~~~(f_+(0)>0)
\eea
We observe that both solutions are smaller than the value $g\approx 1$
obtained by PCAC \cite{Nussinov} or in a
non relativistic quark model (the first
reference in \cite{Yan}). They are also within the present experimental
bound from $D^\star\to D\pi$ decay: $|g|\le 1.7$. In fact from (1.1)
one gets $\Gamma(D^{\star +}\to D^0\pi^+)=g^2 p_\pi^3/(6\pi f^2)=0.20 g^2~
MeV$ and experimentally \cite{PDG} one has  $\Gamma(D^{\star +}\to D^0\pi^+)
<0.6~MeV$.

We can now use the result (2.11) to get information on the coupling $f_+(0)$
for ${\bar B}^0\to\pi^+ e^-{\bar\nu}_e$; we get the values
\bea
f_+^{B\to\pi}(0) &=& -0.67\pm0.20~~~~~~(g=0.48\pm 0.15)\nn\\
f_+^{B\to\pi}(0) &=& 1.08\pm0.20~~~~~~(g=-0.80\pm 0.15)
\eea
for the form factors at $q^2=0$; using again (2.9) for the $q^2$ dependence
(with $m_1=m_{B^\star}=5325~MeV$) we obtain the following predictions for the
branching ratio  (we use $\tau_B=1.24~{\rm psec.}$, eq. (2.10)
with the parameters $\Delta=46~MeV$, $f_B=180~MeV$ and the substitutions
$m_D\to m_B$, $m_{D^\star}\to m_{B^\star}$):
\bea
BR({\bar B}^0\to \pi^+ e^-{\bar\nu}_e) &=& 6\times 10^{-4}\left(\frac{V_{ub}}
{0.004}\right)^2~~~~~~~~~~~~(g=0.48\pm0.15)\\
BR({\bar B}^0\to \pi^+ e^-{\bar\nu}_e) &=& 15.6\times
10^{-4}\left(\frac{V_{ub}}
{0.004}\right)^2~~~~~~(g=-0.80\pm0.15)
\eea
The present experimental bound \cite{PDG}
$BR(B^+\to \pi^0 e^+\nu_e)<2.2\times 10^{-3}$ does not allow for a choice
between (2.13) and (2.14) and we should wait for more precise data. We can
nevertheless get some hints from the decay $D^0\to K^- e^+\nu_e$. In
this case we would get from (2.10)  (with $m_\pi\to m_K$, $m_{D^\star}\to
m_{D_s^\star}=2110~Mev$ and $\Delta=245~MeV$):
\bea
f_+(0)&=& -0.83\pm 0.12~~~~~~~~~~(g=0.48\pm 0.15)\nn\\
f_+(0)&=& 0.21\pm 0.12~~~~~~~~~~(g=-0.80\pm 0.15)
\eea
to be compared to the value obtained from the semileptonic width \cite{PDG}
\be
|f_+(0)|=0.74\pm 0.03
\ee
Eqs. (2.15) and (2.16) point towards a smaller value of $g$, i.e. $g\approx
0.48$, but it should be noted that for $D\to K$ decays the
role of terms with higher derivatives is expected to be not negligible.

\resection{Conclusions}
The calculations we have performed neglect a number of corrections (in
$1/m_Q$, in higher derivatives, non-polar contributions, etc.) so that one
would have to be careful before taking them with an absolute faith.
Nevertheless it seems that
the chiral perturbation theory for mesons
containing a heavy quark, supplemented by information on the leptonic decay
constant from QCD sum rules, can be used to get two possible values for
the $D^\star D\pi$ coupling constant from semileptonic $D\to\pi$ decay.
A (more debatable) application to
$D\to K$ semileptonic decay would favour the smaller value for $g$
($g\approx 0.48$) in the lagrangian of eq. (1.1).

\vspace{1cm}
\noindent
{\bf ACKNOWLEDGMENT} We would like to thank S. De Curtis and D. Dominici
for many interesting discussions.

\end{document}